\documentclass[prb,letterpaper,aps,floatfix,superscriptaddress,twocolumn]{revtex4-1}
\usepackage{graphicx}
\usepackage{amsmath}
\usepackage{subfigure}
\newcommand{\beq}{\begin{equation}}
\newcommand{\eeq}{\end{equation}}
\newcommand{\bk}{{{\bf{k}}}}

\newcommand{\bA}{{\bf{A}}}

\newcommand{\beqa}{\begin{eqnarray}}
\newcommand{\eeqa}{\end{eqnarray}}
\newcommand{\pdg}{{\vphantom \dag}}
\newcommand{\dg}{{\dag}} 
\begin{document}
\title{Weyl fermions and the anomalous Hall effect in metallic ferromagnets}
\author{Y. Chen}
\affiliation{Department of Physics and Astronomy, University of Waterloo, Waterloo, Ontario 
N2L 3G1, Canada}
\author{D.L. Bergman}
\affiliation{Department of Physics, California Institute of Technology, 1200 E. California Blvd, MC114-36, Pasadena, California 91125, USA}
\author{A.A. Burkov}
\affiliation{Department of Physics and Astronomy, University of Waterloo, Waterloo, Ontario 
N2L 3G1, Canada} 
\date{\today}
\begin{abstract}
We reconsider the problem of the anomalous Hall effect in ferromagnetic SrRuO$_3$, incorporating insights 
from the recently developed theory of Weyl semimetals.    
We demonstrate that SrRuO$_3$ possesses a large number of Weyl nodes, separated in momentum space, in its bandstructure. 
While the nodes normally do not coincide with the Fermi energy, unless the material is doped, we show that even the nodes 
inside the Fermi sea have a significant effect on the physical properties of the material. In particular, we show that the 
common belief that (non-quantized part of) the intrinsic anomalous Hall conductivity of a ferromagnetic metal is entirely a Fermi surface 
property, is incorrect: there generally exists a contribution to the anomalous Hall conductivity that arises from topological Fermi-arc surface states, associated with the Weyl nodes, 
which is of the same order of magnitude as the Fermi surface contribution. 
 \end{abstract}
\maketitle
\section{Introduction}
Understanding topological properties of the electronic structure of materials and their experimentally observable consequences has become one of the most 
important themes of the modern condensed matter physics, gradually replacing the traditional focus on the properties, determined by symmetry. 
This shift has become particularly obvious in recent years, after the discovery of topological insulators.~\cite{TI}
An interesting new development in this field is the realization that to possess topologically nontrivial properties 
a material does not have to be an insulator, which significantly expands the range of potential realizations of 
such materials.~\cite{Volovik,Wan11,Ran11,Burkov11,Xu11,Spivak}
In particular, it is now understood that nondegenerate accidental band crossings, named Weyl nodes in Ref.~\onlinecite{Wan11}, which occur generically in any 
three-dimensional material with a broken time reversal or inversion symmetry, have topologically 
nontrivial properties. These crossings act as ``magnetic monopoles" in momentum space and the corresponding 
quantized topological charge gives the band-crossing points stability to perturbations. 

An alternative view of the Weyl nodes, which applies specifically to ferromagnetic materials, in which the nodes appear due to broken 
time reversal symmetry, is based on regarding the  three-dimensional (3D) bandstructure 
as a set of two-dimensional (2D) bandstructures, parametrized by the crystal momentum component $k_z$, 
where the $z$-axis is taken to be along the magnetization direction. 
Weyl nodes in this picture correspond to gap-closing quantum phase transitions, at which Chern numbers of the 2D
bands change pairwise by plus and minus the topological charge of the corresponding node. 
In other words, a Weyl node can be viewed as a quantum Hall transition point in momentum space. 
One interesting consequence, that immediately follows from this, is the existence of chiral Fermi-arc 
surface states, corresponding to pairs of bands with nonzero Chern numbers, where the Fermi arc connects projections of the Weyl nodes with opposite topological charge on the 
sample surface Brillouin zone (BZ).~\cite{Wan11,Burkov11} 

As we show in this paper, this viewpoint is particularly useful in discussing the role of the Weyl nodes in intrinsic anomalous Hall effect (AHE) in metallic ferromagnets.
AHE in ferromagnets is an old problem,~\cite{Karplus54} the interest in which has been revived recently with the realization that topological properties 
of the electronic structure likely play a very important role in it.~\cite{Niu99,MacDonald02,Nagaosa02,Fang03,MacDonald04,Haldane04,Ong04,Nagaosa10}
In particular, here we demonstrate that Weyl nodes, and the associated topological Fermi-arc surface states contribute significantly to the intrinsic anomalous Hall 
conductivity of ferromagnetic metals, which implies that the anomalous Hall conductivity in ferromagnets can not be viewed
as a Fermi surface property. 

\section{Tight-binding model of ${\bf SrRuO_3}$}
While our results likely apply to most metallic ferromagnets, we choose SrRuO$_3$ as the specific material we focus on. 
The main reason for this choice is a relative simplicity of its relevant electronic structure near the Fermi energy, which to a good approximation 
consists of six bands, corresponding to the three $t_{2 g}$ Ru d-orbitals,~\cite{Mazin} which are $2/3$-filled in the undoped material. 
We also assume undistorted cubic perovskite crystal structure for SrRuO$_3$, which should be a good approximation, especially since 
we are interested in topological properties of the bandstructure, insensitive to minor variations of the parameters.

We start from a 6-orbital tight-binding model for SrRuO$_3$, which can be written down based on symmetry considerations.~\cite{Kee12}
The momentum space Hamiltonian has the following form:
\beq
\label{eq:1}
H = \sum_\bk \left[\epsilon^a_{\bk \sigma} \delta_{ab} \delta_{\sigma \sigma'} + f^{a b}_{\bk} \delta_{\sigma \sigma'} 
+ i \lambda \epsilon^{a b c} \tau^c_{\sigma \sigma'} \right] d^\dg_{\bk a \sigma} d^\pdg_{\bk b \sigma'},
\eeq
where summation over repeated orbital and spin indices is implicit. 
The first term in Eq.~\eqref{eq:1} corresponds to spin-split unhybridized $t_{2 g}$-orbital band dispersions:
\beqa
\label{eq:2}
\epsilon^{1 = yz}_{\bk \sigma}&=&- 2 t_1 [\cos(k_y) + \cos(k_z)] - 2 t_2 \cos(k_x) - m \tau^z_{\sigma \sigma}, \nonumber \\
\epsilon^{2 = xz}_{\bk \sigma}&=&- 2 t_1 [\cos(k_x) + \cos(k_z)] - 2 t_2 \cos(k_y) - m \tau^z_{\sigma \sigma}, \nonumber \\
\epsilon^{3 = xy}_{\bk \sigma}&=&- 2 t_1 [\cos(k_x) + \cos(k_y)] - 2 t_2 \cos(k_z) - m \tau^z_{\sigma \sigma}, \nonumber \\
\eeqa
where $t_1$ and $t_2$ are the in-plane and out-of-plane orbital hopping matrix elements, $m$ is the rigid momentum-independent 
exchange spin splitting, which is nonzero in the ferromagnetic phase, and crystal momentum is measured in units of the cubic 
structure lattice constant. 
Second term corresponds to inter-orbital spin-independent hopping, which is induced by hybridization between the oxygen p-orbitals 
and ruthenium d-orbitals:
\beq
\label{eq:3}
f^{ab}_\bk = - 4 f \sin(k_a) \sin(k_b).
\eeq
Finally, the third term corresponds to spin-orbit (SO) interactions, projected onto the $t_{2 g}$ orbital manifold, where 
$\lambda$ denotes the strength of the SO interactions, $\epsilon^{a b c}$ is the fully antisymmetric tensor and 
$\tau^{1,2,3}$ is the triplet of Pauli matrices. 
We should point out here that we will not attempt any quantitative comparison with experiments in this paper and treat parameters 
in Eq.~\eqref{eq:1} with some freedom. What we are after in this work is qualitative features of the bandstructure and their role 
in the AHE, and we expect our results to apply quite generally to metallic ferromagnets, not just to SrRuO$_3$. 

In the absence of the SO interactions, the eigenstates of $H$ exhibit multiple band crossings, which arise due to the spin splitting. 
One might think that once the SO interactions are turned on, these band crossings will be eliminated and replaced by 
gaps of magnitude, proportional to the SO interaction strength $\lambda$ for small $\lambda$.~\cite{Nagaosa10}
Recent work on Weyl semimetals, however, has demonstrated that this is not the case: nondegenerate band crossings 
in three dimensions possess topological stability and can not be eliminated by small perturbations. 
We find that a large number of band crossings in SrRuO$_3$ survive even in the presence of the SO interactions and 
have a very significant qualitative effect on the anomalous Hall conductivity, as will be shown below. 
\begin{figure}[t]
 \centering
 \includegraphics[width=8cm]{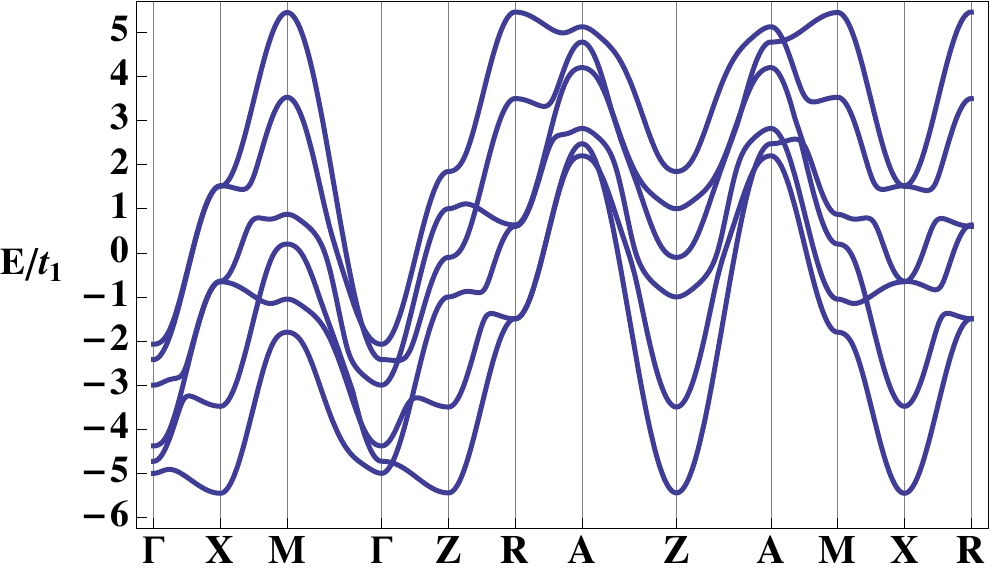}
 \caption{(Color online). Band dispersion for $t_2/t_1 = - 0.2,  f / t_1 = 0.2, \lambda/t_1 = 0.4, m/t_1 = 1$, plotted along high-symmetry directions using tetragonal 
 lattice system notation. There are multiple band crossings between different bands. Some of the crossings 
 are in fact away from high-symmetry directions and can not be seen in this plot.}
 \label{fig:2}
\end{figure}
\section{Intrinsic anomalous Hall effect and Weyl nodes: general discussion}
As is well-known, intrinsic anomalous Hall conductivity can be calculated as an integral of the Berry curvature over all the 
occupied states:~\cite{Nagaosa10}
\beq
\label{eq:4}
\sigma_{xy} = \frac{e^2}{\hbar} \sum_{n}  \int \frac{d^3 k}{(2 \pi)^3} n_F(\epsilon_{n \bk}) \Omega^z_{n \bk},
\eeq
where $n_F(\epsilon)$ is the Fermi-Dirac distribution function, $\epsilon_{n \bk}$ is the band dispersion, 
and ${\boldsymbol \Omega}_{n \bk}$ is the Berry curvature vector, which is a curl of the Berry connection 
$\bA_{n \bk}=i \langle n \bk |{\boldsymbol \nabla}_k | n \bk \rangle$, and may be thought of as an analog of magnetic 
field in momentum space. 
One way to understand the important role, played by the Weyl nodes in the intrinsic AHE, is to realize that these nodes act as 
magnetic-monopole-like sources of the Berry curvature field, which is divergenceless in the absence of such sources.
For our purposes, however, it will be more useful to adopt a different viewpoint. 
Namely, since the magnetization $m$ along the $z$-axis introduces a preferred direction and reduces the 
cubic symmetry down to tetragonal, we can separate the 
3D momentum space integration in Eq.~\eqref{eq:4} into a 1D integral over $k_z$ and 
a 2D integral over $\bk_{\perp} = (k_x, k_y)$:
\beq
\label{eq:5}
\sigma_{xy} = \int_{-\pi}^{\pi} \frac{d k_z}{2 \pi} \sigma^{2D}_{xy}(k_z), 
\eeq
where
\beq
\label{eq:6}
\sigma^{2D}_{xy}(k_z) = \frac{e^2}{\hbar} \sum_n \int \frac{d^2 k_{\perp}}{(2 \pi)^2} n_F[\epsilon_{n \bk_{\perp}}(k_z)] \Omega^z_{n \bk_{\perp}}(k_z). 
\eeq
We may then regard $\sigma^{2D}_{xy}(k_z)$ as the Hall conductivity of a set of 2D systems, parametrized by $k_z$. 
This representation of $\sigma_{xy}$ turns out to be extremely useful, as we show below.
\begin{figure}
\subfigure[]{
   \label{fig:3a}
  \includegraphics[width=7cm]{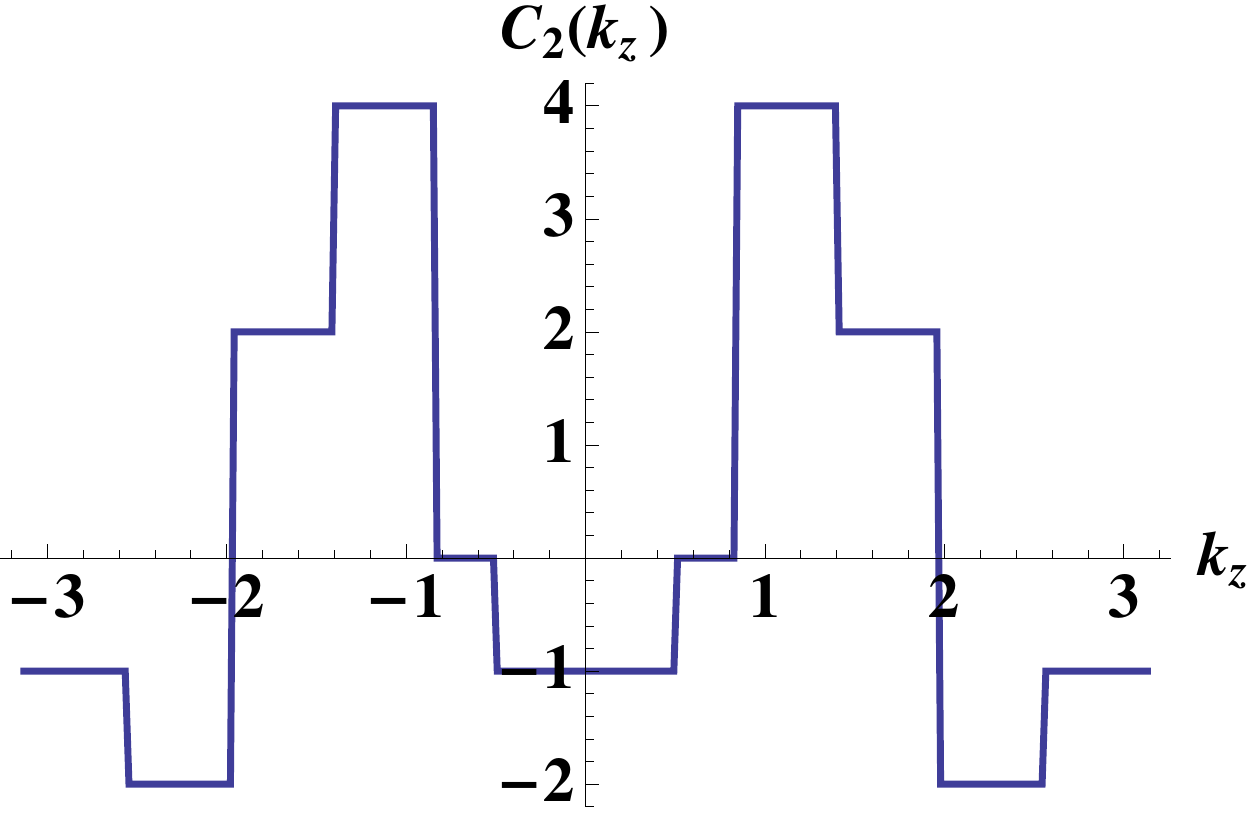}}
\subfigure[]{
  \label{fig:3b}
   \includegraphics[width=7cm]{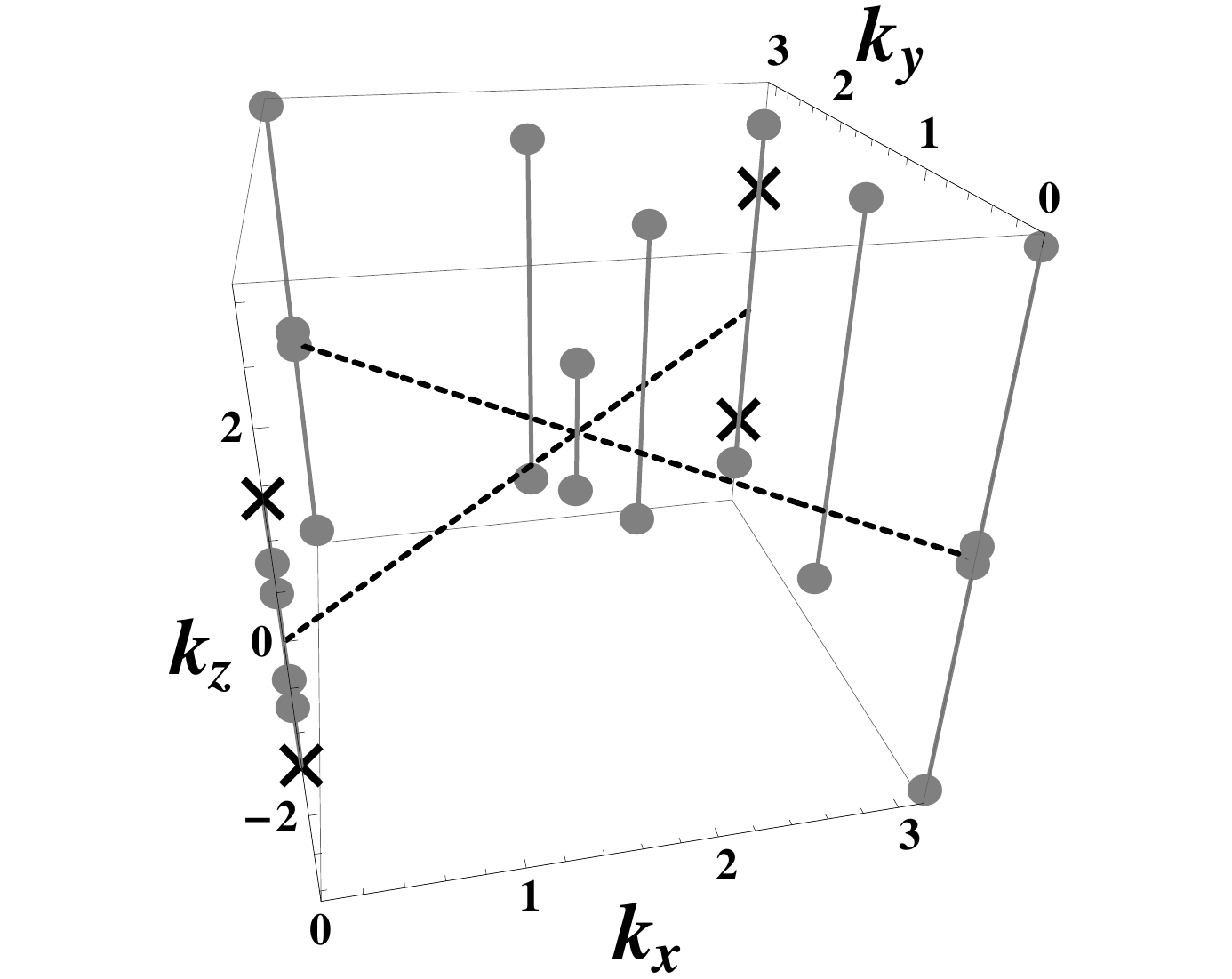}}
  \caption{(Color online). (a) Chern number for the second lowest band as a function of $k_z$ for the same parameter values as in Fig.~\ref{fig:2}. The Chern number experiences multiple 
  integer-valued jumps as a function of $k_z$ in the first BZ. The jumps are the evidence of the existence of Weyl nodes. (b) Locations of all Weyl nodes in the bandstructure of SrRuO$_3$. 
  Only one quarter of the first Brillouin zone is shown. The remaining points are obtained by successive $\pi/2$-rotations around the $z$-axis. Dots are linear (charge one) Weyl nodes, 
  while crosses denote quadratic (charge two) Weyl nodes. The solid (connecting pairs of nodes with opposite topological charge) and dashed lines are guides to the eye.} 
  \label{fig:3}
\end{figure}

As was first pointed out by Haldane,~\cite{Haldane04} we can use Stokes theorem and rewrite $\sigma^{2D}_{xy}$ in the following way:
\beq
\label{eq:7}
\sigma^{2D}_{xy}(k_z) = \frac{e^2}{2 \pi h} \sum_n \oint d {\bk} \cdot \bA_{n \bk}(k_z), 
\eeq
where the integral is along the 1D Fermi surface of a 2D system, corresponding to a given $k_z$.   
This suggests that the intrinsic anomalous Hall conductivity may be thought of as a Fermi surface property, like all other 
transport properties of metals.~\cite{Haldane04}
However, there is a subtlety in this argument. The Stokes theorem applies only when the corresponding band has a zero Chern number.~\cite{Thouless82,Hatsugai93}
Chern number is an {\em obstruction} to the Stokes theorem~\cite{Hatsugai93} and the correct form of Eq.~\eqref{eq:7} is:
\beq
\label{eq:8}
\sigma^{2D}_{xy}(k_z) = \frac{e^2}{2 \pi h} \sum_n \oint d {\bk} \cdot \bA_{n \bk}(k_z) + \frac{e^2}{h} \sum'_n C_n(k_z), 
\eeq
where
\beq
\label{eq:9}
C_n(k_z) = \frac{1}{2 \pi} \int d^2 k_{\perp} \Omega^z_{n \bk_{\perp}} (k_z),
\eeq
is the Chern number of the two-dimensional band $n$ at momentum $k_z$ and the sum over $n$ in the 
second term in Eq.~\eqref{eq:8} is restricted to 
completely filled 2D bands only, which is indicated by the prime. 
To make Eq.~\eqref{eq:8} well-defined we may regard the first BZ as an open square rather than a torus, so that
when a band is completely filled, the line integral in the first term becomes an integral over the BZ boundary and gives the corresponding 
Chern number. Gauge ambiguity in the first term may be eliminated by comparing Eq.~\eqref{eq:8} with the explicitly gauge-invariant 
result, obtained from Eq.~\eqref{eq:6}. 

If Weyl nodes were not present in the bandstructure, all the Chern numbers
$C_n(k_z)$ would not in fact depend on $k_z$, and it is normally assumed that they are 
zero, since a very large spin splitting would be needed to create a band with a constant 
nonzero $C_n(k_z)$, implying a contribution to the total Hall conductivity from this band, 
quantized in units of $e^2 G / 2 \pi h$, where $G =2 \pi$ is the smallest reciprocal lattice vector. 
Thus one may conclude that the anomalous Hall conductivity is entirely a Fermi surface property.~\cite{Haldane04}
This, however, is incorrect, since $C_n(k_z)$ are not independent of $k_z$ in 
the presence of Weyl nodes, and they are in general present in any ferromagnet, even when the magnetization is small.
Indeed, as discussed above, a Weyl node may be thought of as a gap-closing transition 
point for the 2D bandstructure, parametrized by $k_z$. 
At the transition point, the Chern numbers of the two bands, that touch at the Weyl node, 
change by $\pm$ the topological charge of the node.~\cite{Wan11,Ran11,Burkov11,Xu11}
The corresponding contribution to the total 3D Hall conductivity is then not 
quantized in units of $G$, but is still not a Fermi surface property: it instead corresponds to the 
Fermi-arc surface states, associated with pairs of Weyl nodes with opposite topological charge. 
\section{Intrinsic anomalous Hall effect and Weyl nodes: ${\bf SrRuO_3}$}
Let us now see how the above picture is realized in SrRuO$_3$. 
The 6-band tight-binding Hamiltonian of SrRuO$_3$ 
Eq.~\eqref{eq:1} is easily diagonalized numerically. 
The bandstructure, plotted along high-symmetry directions in the first BZ in the 
standard way, is shown in Fig.~\ref{fig:2}. 
Some of the Weyl nodes are in fact clearly visible in Fig.~\ref{fig:2}. 
In general, however, finding all the nodes, along with their topological charges, 
is quite a difficult task.
What makes it significantly easier is the relation between the Weyl nodes and the 
change of the Chern number as a function of $k_z$, discussed above. 
To evaluate the Chern numbers, we use a discrete lattice version of  Eq.~\eqref{eq:9},~\cite{Vanderbilt07} 
corresponding to the discrete values of the crystal momentum in a finite sample with periodic 
boundary conditions.  
Namely we calculate the lattice Berry connection fields as:
\beq
\label{eq:10}
A^n_{\bk, \mu} = \langle n \bk | n \bk + \mu \rangle, 
\eeq
where $\mu$ are the nearest-neighbor vectors of the square momentum-space lattice with a fixed $k_z$. 
The Chern number is then evaluated as:
\beq
\label{eq:11}
C_n(k_z) = \frac{1}{2 \pi} \sum_\bk \textrm{Im} \ln(A^n_{\bk, \hat x} A^n_{\bk + \hat x, \hat y} A^n_{\bk + \hat x +\hat y, -\hat x} A^n_{\bk + \hat y, -\hat y}). 
\eeq   
An example of the resulting plot of $C_n(k_z)$ is shown in Fig.~\ref{fig:3}.
Every discrete jump of the Chern number by an integer is due to the presence of 
Weyl nodes in the plane in 3D BZ, corresponding to the given value
of $k_z$. The magnitude of the jump is equal to the total topological charge of Weyl nodes 
in a given plane. The information obtained from these plots, plus symmetry considerations, confirmed 
by an explicit examination of the band dispersions, allow us to identify all the Weyl nodes in the 
bandstructure of SrRuO$_3$, as shown in Fig.~\ref{fig:3}.     
We find a total of 22 pairs of nodes with opposite topological charges in the first BZ. 
Two of those pairs, separated along the $(k_x, k_y)= (0,0)$ and $(k_x, k_y) = (\pi, \pi)$ lines, 
correspond to double-Weyl nodes with topological charges $\pm 2$. These are stabilized 
by the cubic symmetry of the undistorted perovskite structure and will be split each into two 
pairs of ordinary Weyl nodes with single unit of topological charge when the orthorhombic distortion, 
typically present in real SrRuO$_3$ material, is taken into account.~\cite{Bernevig12} 
The rest of the nodes have topological charge $\pm 1$. 
Of those, four pairs are located on the same lines as the double-Weyl nodes, four pairs 
are on the $(k_x, k_y) = (0, \pi), (\pi, 0)$ lines, eight pairs are located in the $k_x = \pm k_y$ planes,  
and four pairs are located in the $k_x = \pi$ and $k_y = \pi$ planes. 
The nodes are generally not at the same energy, unless required by symmetry. 
 \begin{figure}[t]
 \subfigure[]{
  \label{fig:4a}
   \includegraphics[width=7cm]{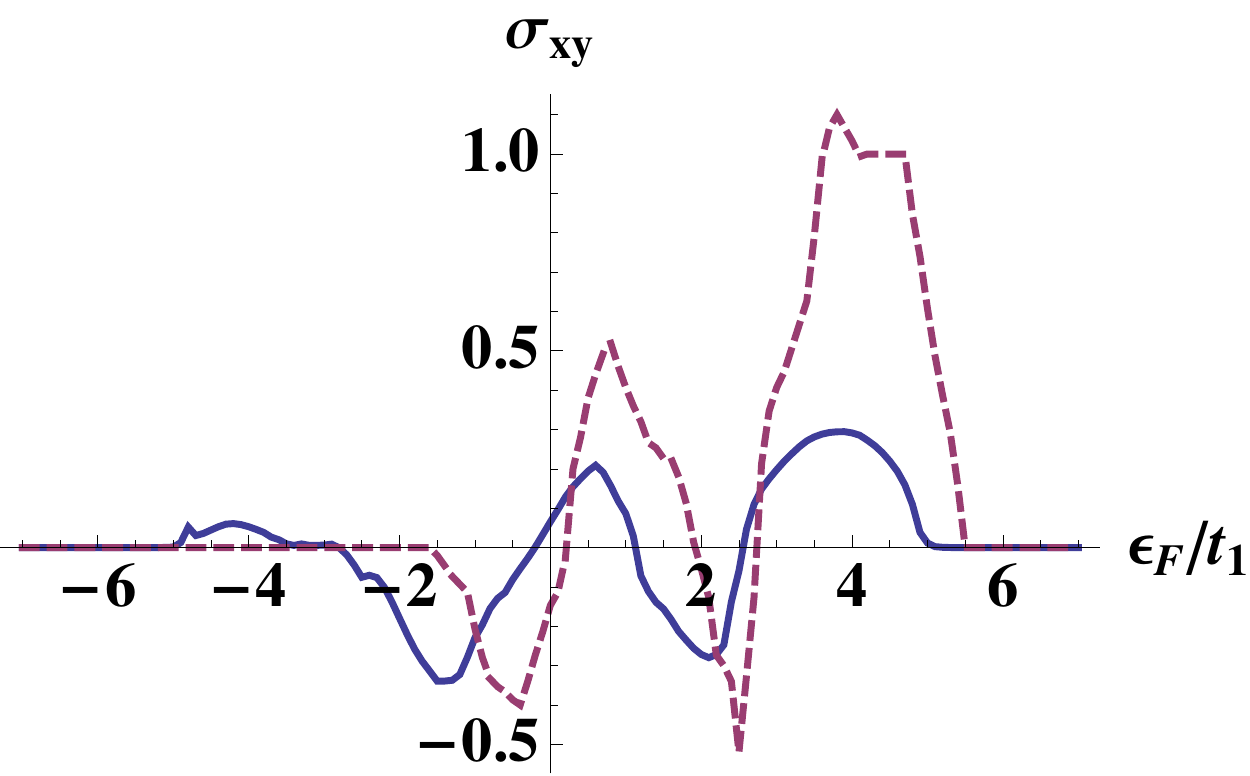}}
  \subfigure[]{
  \label{fig:4b}
  \includegraphics[width=7cm]{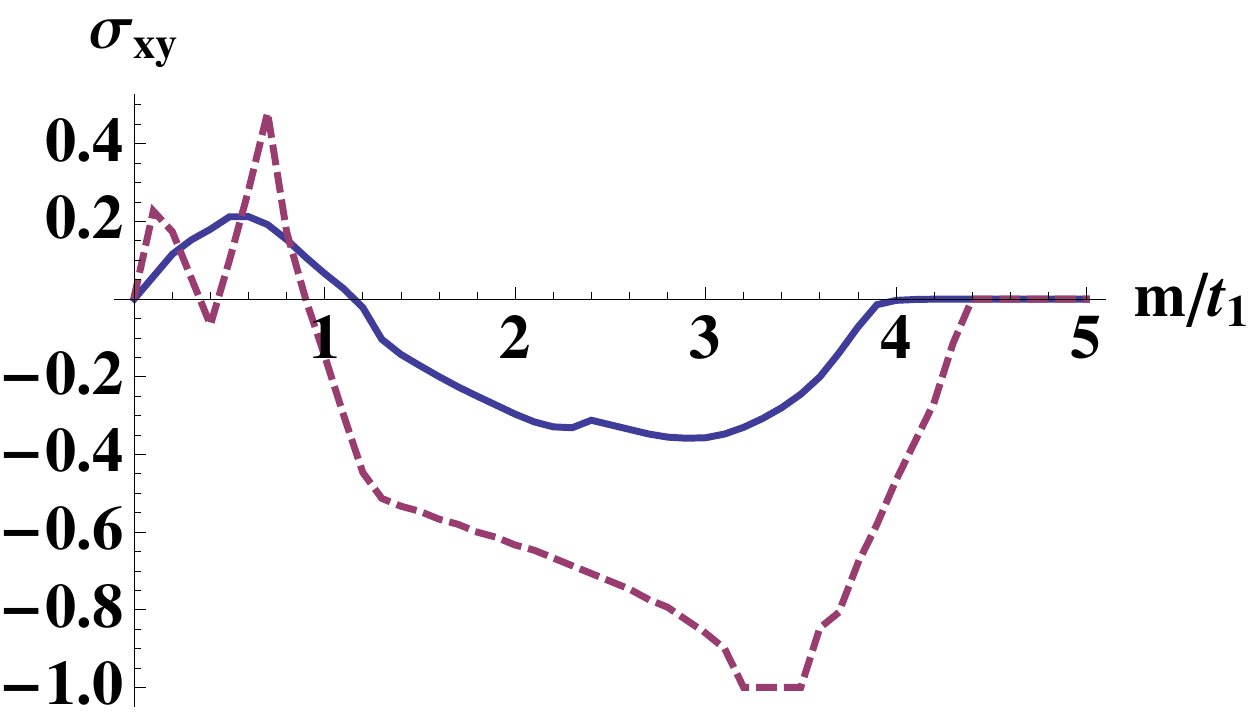}}
    \caption{(Color online). Total anomalous Hall conductivity $\sigma_{xy}$ (solid line) and the edge-state contribution to the conductivity $\sigma^{edge}_{xy}$ (dashed line) as a function of Fermi energy 
    $\epsilon_F$ for $m/t_1 = 1$ (a), and magnetization $m$ for $\epsilon_F/t_1 = 0$ (b).  $\sigma_{xy}$ is in units of $e^2/ h$ (the lattice constant 
    is set to unity).}
     \label{fig:4}
\end{figure}

\begin{figure}[t]
\includegraphics[width=7cm]{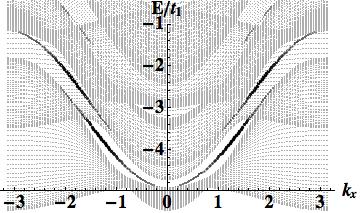}
\caption{Example of a calculated chiral surface state dispersion in the $k_z=0$ plane for a sample with open boundaries, perpendicular to the $y$-axis. 
The intensity of gray is proportional to the degree of surface localization of the corresponding state, measured by the inverse participation 
ratio of its wavefunction. The parameter values are the same as in Fig.~\ref{fig:2}. Two pairs of chiral surface states, corresponding to the double-Weyl nodes, 
separated along the $(k_x, k_y) = (0, 0)$ line, is clearly visible. Each pair of chiral surface states is localized on one of the sample surfaces.}
\label{fig:5}
\end{figure}

We can now explicitly identify two distinct contributions to the total anomalous Hall conductivity
of SrRuO$_3$: the contribution arising from the Femi-arc surface states and the contribution, associated 
with the bulk Fermi surface. 
We evaluate the surface state part of the Hall conductivity by summing the Chern-number contributions of all completely 
filled 2D bands at every $k_z$:~\cite{Hatsugai93}
\beq
\label{eq:12}
\sigma^{edge}_{xy} =  \frac{e^2}{h} \int_{-\pi}^{\pi} \frac{d k_z}{2 \pi} \sum'_n C_n(k_z).
\eeq 
The bulk Fermi surface contribution is then the difference between the total Hall conductivity, given by Eq.~\eqref{eq:4},
and the surface state contribution, Eq.~\eqref{eq:12}. 
The total anomalous Hall conductivity is evaluated by summing the lattice Berry curvature, Eq.~\eqref{eq:11}, 
over all states below the Fermi energy.  
The results are shown in Fig.~\ref{fig:4}. It is seen that $\sigma_{xy}$ is a highly nonmonotonic function of both 
the Fermi energy (which can be varied to some degree by doping) and the magnetization.
The peaks and dips in the dependence of $\sigma_{xy}$ on $\epsilon_F$ correspond to the Fermi level passing 
through the Weyl nodes. 
It can also be seen that the Fermi-arc surface state contribution is always of the same order as the Fermi surface 
contribution and can not be regarded as an insignificant correction.     
An example of a chiral surface state dispersion, calculated for a sample with open boundaries, is shown in 
Fig.~\ref{fig:5}.  

Eq.~\eqref{eq:12} appears to suggest that all Weyl nodes below the Fermi energy contribute 
to the anomalous Hall conductivity. 
This, however, is not correct. As discussed above, Weyl nodes may be thought of as points of quantum Hall 
transitions in momentum space, at which the corresponding 2D bands experience equal in magnitude 
but opposite in sign change of the Chern number. 
This means, in particular, that when both 2D bands in the pair, joined by a pair of Weyl nodes, are filled, 
the total contribution of this pair of nodes to the anomalous Hall conductivity is zero. 
\section{Conclusions}
In conclusion, we have identified the role, played by nondegenerate band-touching nodes (Weyl nodes) 
in the intrinsic AHE in ferromagnetic metals. We have shown that, in general, 
even the non-quantized part of the anomalous Hall conductivity is not a Fermi liquid property, in the sense 
that a significant part of it arises from the Femi-arc surface states, associated with pairs of Weyl nodes 
of opposite chirality, and not with bulk states on the Fermi surface. 
It is clear from our analysis that any generic model of intrinsic AHE in ferromagnetic metals must
incorporate Weyl nodes and the currently used models are lacking in this regard. 

It would be very interesting to try to observe chiral edge states in SrRuO$_3$ experimentally. 
One possibility would be to use the ARPES technique, whose usefulness in this regard has been 
clearly demonstrated in the study of topological insulators.~\cite{TI} 
Another interesting possibility is the scanning SQUID susceptometry, which can directly 
image the edge currents.~\cite{Moler} 

\begin{acknowledgments}
We acknowledge useful discussions with L. Balents, H.-Y. Kee, Y.-B. Kim, R. Lutchyn, and A.H. MacDonald. 
Financial support was provided by NSERC of Canada and a University of Waterloo start up grant. 
AAB acknowledges the hospitality of the Aspen Center for Physics, funded by NSF grant PHY-1066293.   
\end{acknowledgments}

\end{document}